\documentclass[prb,twocolumn,showpacs]{revtex4}
\usepackage{graphicx}

\begin{document}

\title{Low-temperature electrical transport in bilayer manganite
La$_{1.2}$Sr$_{1.8}$Mn$_{2}$O$_{7}$ }
\author{C. L. Zhang, X. J. Chen, and C. C. Almasan}
\affiliation{Department of Physics, Kent State University, Kent OH 44242,USA}
\author{J. S. Gardner}
\affiliation{Chalk River
Laboratory, Chalk River ON KOJ 1PO, Canada}
\author{J. L. Sarrao}
\affiliation{Los Alamos National Laboratory, Los Alamos, NM 87545,USA}
\date{Received 17 October 2001}

\begin{abstract}
The temperature $T$ and magnetic field $H$ dependence of anisotropic 
in-plane $\rho_{ab}$ and
out-of-plane $\rho_{c}$  resistivities have been investigated in 
single crystals of the
bilayer manganite  La$_{1.2}$Sr$_{1.8}$Mn$_{2}$O$_{7}$. Below the
Curie transition temperature $T_c=$ 125 K, $\rho_{ab}$ and $\rho_{c}$ 
display almost the
same temperature dependence with an up-turn around 50 K. In the 
metallic regime (50 K $\leq T
\leq$ 110 K), both $\rho_{ab}(T)$ and $\rho_{c}(T)$ follow a
$T^{9/2}$ dependence, consistent with the two-magnon
scattering. We found that the value of the proportionality coefficient
$B_{ab}^{fit}$ and the ratio of the exchange interaction $J_{ab}/J_c$ 
obtained by fitting
the data are in excellent agreement with the calculated $B_{ab}$ 
based on the two-magnon
model and
$J_{ab}/J_c$ deduced from neutron scattering, respectively. This 
provides further support
for this scattering mechanism. At even lower
$T$, in the non-metallic regime ($T<$ 50 K), {\it both} the in-plane
$\sigma_{ab}$ and  out-of-plane $\sigma_{c}$ conductivities obey a 
$T^{1/2}$ dependence,
consistent with weak localization
effects.  Hence, this demonstrates the three-dimensional
metallic nature of the bilayer manganite 
La$_{1.2}$Sr$_{1.8}$Mn$_{2}$O$_{7}$ at $T<T_c$.
\end{abstract}

\pacs{75.30.Vn, 72.15.-v, 72.10.-d}

\maketitle

\section{Introduction}

There is growing interest in the low-temperature electrical transport
phenomena of perovskite manganites in order to elucidate the
microscopic origin of the colossal magnetoresistance (CMR)
effect.\cite{schi,jaim,okud,okud2,zhao1,zhao2} Recent studies on
single
crystals of the three-dimensional 3D pseudocubic compounds 
R$_{1-x}$A$_{x}$MnO$_{3}$
($R$ is a trivalent rare-earth ion and $A$ is a divalent
alkaline-earth ion) show a $T^{2}$ dependence of the
resistivity,\cite{jaim,okud,okud2} which has been interpreted as
either
electron-electron \cite{okud,okud2} or one-magnon \cite{jaim}
scattering. However, low-temperature resistivity measurements of
epitaxial thin films of La$_{1-x}$Ca$_{x}$MnO$_{3}$ provide support
for the presence of small-polaron conduction in the
ferromagnetic (FM) state.\cite{zhao1,zhao2} In the same system, a
$T^{9/2}$ term attributed to electron-magnon scattering was also
found in the $T$ dependence of resistivity at
low-temperatures.\cite{schi,zhao2} Therefore, the low-temperature
electrical
transport mechanism of manganites remains controversial and is far
from being fully understood.

The bilayer manganite La$_{2-2x}$Sr$_{1+2x}$Mn$_{2}$O$_{7}$  has
proven to be a fruitful system for understanding the CMR and has
become the focus of many recent investigations.\cite{mori,kimu} This
is an ideal system for the study of the low-temperature
conduction mechanism. Its reduced dimensionality gives rise to
anisotropic characteristics of charge transport and magnetic
properties and also enhances the CMR effect near the magnetic transition
temperature, although at the cost of reducing it to about 100 K.\cite{mori}
The magnetic structure of heavily
doped  bilayer compounds always shows the coexistence of FM and
antiferromagnetic (AFM) correlations.\cite{perr,hiro,kubo1,osbo} The
magnetic correlations are predominantly FM within the
two-dimensional MnO$_{2}$ layers, while the magnetic coupling between
the MnO$_{2}$ layers changes from FM for $x \leq$ 0.4 to
canted AFM for $x>$ 0.4.  The interplay between FM double exchange
and AFM superexchange interactions between Mn ions in these
compounds becomes more subtle and is expected to be responsible for
the unusual transport properties observed in the bilayer
manganites. For examples, (i) an up-turn in the resistivity at low
temperatures is generally observed for the $x=$ 0.30,\cite{qing2}
0.35,\cite{okuda} 0.38,\cite{qing2} and 0.4\cite{mori,qing1,chun}
samples; (ii) despite the anisotropic crystal
structure,\cite{mitc} $\rho_{ab}$ and $\rho_{c}$ of the $x=$ 0.38
compound display virtually identical temperature dependences at
low-temperatures, indicating the same conduction mechanism in both
$ab$ and $c$ directions.\cite{qing2}

A lot of effort has been devoted to understanding the low-temperature
electrical transport properties of bilayer manganites. Okuda,
Kimura, and Tokura \cite{okuda} found that the in-plane conductivity
$\sigma_{ab}$ of $x=$ 0.35 single crystals is almost
proportional to $T^{1/2}$ for $T<$ 4 K, indicating weak localization.
This square-root temperature dependence of $\sigma_{ab}$ was
also observed in the $x=$ 0.40 samples.\cite{chun} Recently,
Abrikosov, based on the theory of quantum interference, has shown that
$\sigma_{ab}(T)$ and $\sigma_{c}(T)$ should be isotropic and
proportional to $\tau_{\varphi}^{-1/2}$  (the phase coherence
destruction probability $\tau_{\varphi}^{-1}\propto T$ for the 3D 
conduction) in
the 3D (low-temperature) regime.\cite{abri} It is,
therefore, desirable from the experimental point of view to determine
$\sigma_{c}(T)$ of bilayer manganites and, hence, to establish which
scattering mechanism is mainly responsible for the low-temperature
transport properties of this system. An important question is
also whether there is any common conduction mechanism responsible for
the low-temperature electrical transport in both the
infinite-layer and bilayer manganites.

In this paper, we address the above issues through resistivity
measurements on La$_{1.2}$Sr$_{1.8}$Mn$_{2}$O$_{7}$ single crystals,
performed at low temperatures and in magnetic fields applied parallel to
the $ab$ plane. The anisotropic resistivities $\rho_{ab}(T)$
and $\rho_{c}(T)$ are proportional to $T^{9/2}$ in the
intermediate temperature regime below $T_{c}$. This points
toward a two-magnon scattering mechanism responsible for the
electrical dissipation in this $T$ range. The validity of this 
scattering mechanism is further
supported by the $H$ dependence of the proportionality coefficients 
$B_{ab,c}^{fit}$
obtained by fitting the data and by the fact that the value of
$B_{ab}^{fit}$ and the ratio of the exchange interaction $J_{ab}/J_c$ 
obtained by fitting
the data are in excellent agreement with the calculated $B_{ab}$ 
based on the two-magnon
model and
$J_{ab}/J_c$ deduced from neutron scattering, respectively. At lower 
temperatures, in the
non-metallic regime ($\partial
\rho/
\partial T<0$), {\it both} anisotropic  conductivities 
$\sigma_{ab}(T)$ and $\sigma_{c}(T)$
exhibit a $T^{1/2}$
dependence, a result of electron-electron correlations, which is
consistent with the weak localization effect in a
3D disordered metal. The common mechanisms
responsible for both the in-plane and out-of-plane electrical transport
indicate the 3D metallic nature of the bilayer manganite 
La$_{1.2}$Sr$_{1.8}$Mn$_{2}$O$_{7}$
at $T<T_c$.

\section{Experimental Details}

Single crystals of La$_{1.2}$Sr$_{1.8}$Mn$_{2}$O$_{7}$ were grown from
sintered rods of the same nominal composition by the
floating zone method using a mirror furnace, as described in detail
elsewhere.\cite{moreno} Plate-shaped crystals were separated
mechanically from the bar. X-ray diffraction confirmed that the surfaces
of the plates are parallel to the crystallographic $ab$
plane. The crystal chosen for systematic transport measurements had
mirror surfaces on both faces. We determined $\rho_{ab}$ and
$\rho_{c}$ as functions of temperature $T$ ($1.9\leq T \leq400$ K) and
magnetic field $H$ ($0\leq H \leq14$ T) by performing
multi-terminal transport measurements using the electrical contact
configuration of the flux transformer, as described
previously.\cite{levin1,levin2} The magnetic field $H$ was applied
parallel to the MnO$_{2}$ layers ($H\parallel ab$ plane).

\section{Results and Discussion}

The $T$ dependence of the magnetic susceptibility $\chi$ measured in
a low $H$ ($H=10$ Oe) is shown in the inset to Fig. 1(b). The
sample exhibits a paramagnetic (PM) - FM transition at the Curie transition
temperature $T_{c}=$ 125 K, which is consistent with  previous
reports.\cite{mori,mitc} At $T<$  50 K, the susceptibility slightly decreases
with decreasing
$T$, in good agreement with recent reports.\cite{chun,mitc} As
typically seen in other studies,\cite{mori,mitc,pott} an
additional transition appears around 290 K, most likely due to trace
amounts of impurities  \cite{oser} (intergrowth) that, however, represent
only about 0.1 \% of the volume fraction of the sample.\cite{pott}

\begin{figure}[t]
\begin{center}
\includegraphics[width=\columnwidth]{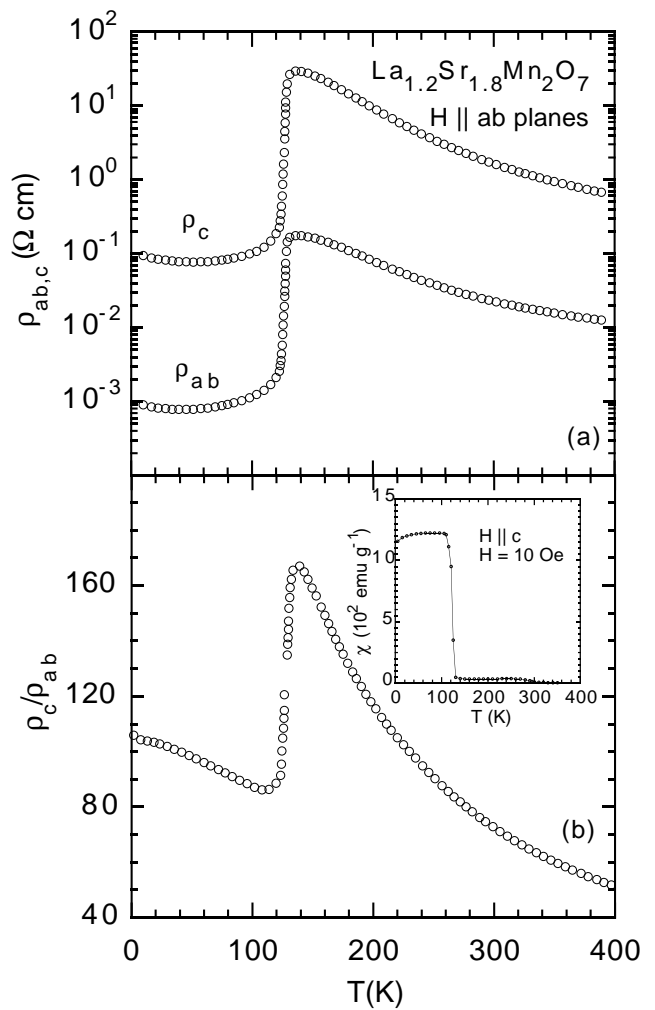}
\end{center}
\caption{ Temperature $T$ dependences of (a) in-plane $\rho_{ab}$ and
out-of-plane $\rho_{c}$ resistivities measured in zero magnetic field
and (b) anisotropy $\rho_{c}/\rho_{ab}$ for
La$_{1.2}$Sr$_{1.8}$Mn$_{2}$O$_{7}$. Inset: $ac$ susceptibility $\chi$ as
a function
of temperature measured in a magnetic field
$H=$ 10 Oe.}
\end{figure}

Figures 1(a) and 1(b) show the temperature profiles of $\rho_{ab}$
and $\rho_{c}$, and  of the anisotropy $\rho_c/\rho_{ab}$,
respectively, measured in zero magnetic field. The metal-insulator transition
takes place at $T_{MI}=$ 130 K for both situations:
current parallel ($\rho_{ab}$) and perpendicular ($\rho_{c}$) to the
MnO$_{2}$ layers. The anisotropy $\rho_{c}/\rho_{ab}$ increases with decreasing
$T$, reaches its maximum value
of 165 at $T_{MI}$, decreases abruptly just below $T_c$, and depends
weakly on $T$ for
$T<T_{c}$ with an average value of $\sim$90, comparable to other
measurements.\cite{mori,qing1} We note that the values of both
$\rho_{ab}$
and $\rho_{c}$ are, over the whole measured $T$ range, appreciably
smaller than those reported previously,\cite{mori} attesting to
the high quality of our single crystal.

The $H$ and $T$ dependences of $\rho_{ab}$ and $\rho_{c}$ are shown
in Figs. 2(a) and 2(b), respectively, for the temperature range
from 2 to 120 K. The characteristic negative magnetoresistivity in both
$\rho_{ab}$ and
$\rho_{c}$ is clearly seen over the whole
$T$ range. At the same time, $\rho_{c}/\rho_{ab}$ has a weaker $H$
dependence [see inset to Fig. 2(b)]. An interesting feature of
$\rho_{ab}(T)$ and $\rho_{c}(T)$ is the appearance of a weak non-metallic
behavior for $T<$ 50 K in zero field, which shifts to higher temperatures with
increasing magnetic field.

\begin{figure}[t]
\begin{center}
\includegraphics[width=\columnwidth]{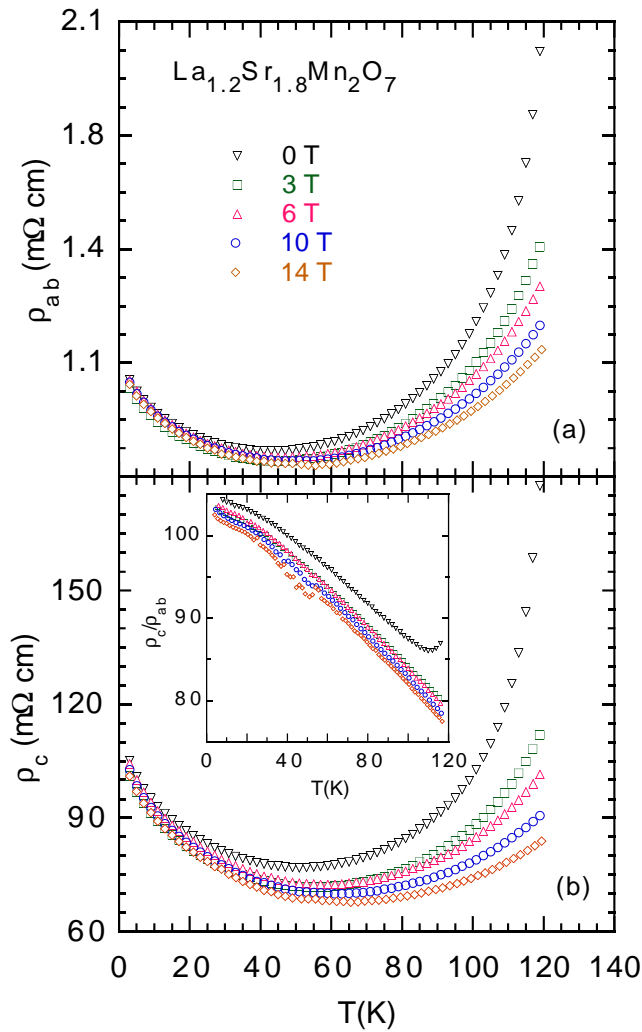}
\end{center}
\caption{Plot of (a) in-plane resistivity $\rho_{ab}$
and (b) out-of-plane resistivity $\rho_{c}$ vs temperature $T$ for
La$_{1.2}$Sr$_{1.8}$Mn$_{2}$O$_{7}$ measured in applied magnetic fields $H$ up
to 14 T. Inset: Anisotropy
$\rho_{c}/\rho_{ab}$ vs $T$ in the temperature range from 2 to 120 K
measured under various $H$.}
\end{figure}

In the infinite-layer La$_{1-x}$Sr$_{x}$MnO$_{3}$ (0.10 $\leq x \leq$
0.17) samples, a similar low-temperature up-turn in
resistivity has been identified with the vestiges of the structural
$O^{\star}$ to $O^{\prime}$ transition in the FM
phase.\cite{dabr} Temperature-dependent neutron diffraction studies
on La$_{1.2}$Sr$_{1.8}$Mn$_{2}$O$_{7}$ do not show any
structural transition occurring at low-temperatures. \cite{mitc} The
up-turn in the resistivity of
La$_{1.2}$Sr$_{1.8}$Mn$_{2}$O$_{7}$ at low temperatures is
accompanied by a decrease of the $ac$ susceptibility, which is related
to the reentrant spin glass phase,\cite{chun} an indication of a
crossover from a FM to a canted state.\cite{genn} The canted
structure at low temperatures comes from the competing interactions
along the $c$-axis Mn-O-Mn bonds, $i.e.,$ the AFM superexchange
interaction between half-filled $t_{2g}$ orbitals and the FM
double-exchange interaction via $e_{g}$ conduction
electrons.\cite{genn}

In ordinary FM metals, the contribution to resistance from the
$s$-$d$ interaction is known to be proportional to $T^{2}$ at low
temperatures.\cite{kasu,good}  This $T^{2}$ dependence of resistivity
has been generally observed in single crystals of
La$_{1-x}$Sr$_{x}$MnO$_{3}$ (x $>$ 0.18),\cite{okud}
La$_{1-x}$Ca$_{x}$MnO$_{3}$ (x $\geq$ 0.22),\cite{okud2} and
La$_{0.67}$(Pb,Ca)$_{0.33}$MnO$_{3}$,\cite{jaim} and has been
attributed to either electron-electron \cite{okud,okud2} or
one-magnon \cite{jaim} scattering.
However, our low-temperature resistivities data (50 K $ \leq T
\leq$ 110 K) of La$_{1.2}$Sr$_{1.8}$Mn$_{2}$O$_{7}$
single crystals measured in various applied magnetic fields do not 
follow a $T^2$ dependence
of the form
$\rho (T)=\rho_{0}+AT^{2}$.
Therefore, the electron-electron scattering mechanism does not contribute to
$\rho_{ab,c}(T)$. This indicates that the conduction  mechanism in 
this bilayer manganite is
different from that in the infinite-layer compounds. In fact, 
specific heat measurements on
bilayer manganites \cite{okuda} show that their reduced
dimensionality, compared to infinite-layer manganites, enhances the
magnetic specific heat, but does not affect the electronic
specific heat coefficient $\gamma$, indicating the presence of an
anomalous carrier scattering process such as electron-phonon,
electron-magnon, or a combination of them.

Polaronic transport has been recently shown to be a possible
conduction mechanism in R$_{1-x}$A$_{x}$MnO$_{3}$ at
$T<T_c$.\cite{zhao1,zhao2,alex} Zhao $et$ $al.$\cite{zhao1,zhao2}
reported that the resistivity below 100 K in epitaxial thin films
of La$_{1-x}$Ca$_{x}$MnO$_{3}$ ($x =$ 0.25 and 0.4) grown on (100)
LaAlO$_{3}$ substrates can be well fitted with
$\rho=\rho_{0}+E\omega_{s}/\sinh^{2}(\hbar\omega_{s}/2k_{B}T)$
($\omega_{s}$ is the frequency of a soft
optical mode), providing evidence for small-polaron metallic
conduction in the FM state. We found that our
resistivity data of La$_{1.2}$Sr$_{1.8}$Mn$_{2}$O$_{7}$ can not be
adequately fitted with this expression. Presently,
there is no experimental evidence for the presence of small
polarons in the FM state of bilayer manganites. In
fact, recent Raman scattering data on
La$_{1.2}$Sr$_{1.8}$Mn$_{2}$O$_{7}$ single crystals suggest the
formation of small
polarons {\it only} at $T>T_{c}$.\cite{rome} Moreover, X-ray and
neutron scattering measurements \cite{dolo} on
La$_{1.2}$Sr$_{1.8}$Mn$_{2}$O$_{7}$ directly demonstrate that the
polarons disappear abruptly at the FM transition because of the
sudden charge delocalization. On the other hand, the giant
magnetothermal conductivity \cite{mats} observed in
La$_{1.2}$Sr$_{1.8}$Mn$_{2}$O$_{7}$ indicates that spin-fluctuation
scattering or magnon contribution is dominant over other
carrier scattering processes.

Kubo and Ohata \cite{kubo} calculated the low-temperature resistance
produced by scattering of holes by spin waves, on the basis of
an effective Hamiltonian for double exchange in the spin wave
approximation. They have found that the contribution from the
two-magnon scattering process is proportional to $T^{9/2}$ and that
the proportionality coefficient has, in the case of a simple
parabolic band, the analytical expression given by
\begin{equation}
B=\frac{3\hbar R^{6} k_{F}^{5} }{ 32 \pi e^{2} S^{2}}
(\frac{m}{M})^{9/2}(\frac{k_{B}}{E_{F}})^{9/2}
(2.52+0.0017\frac{M}{m}).
\end{equation}
Here $R$ is the hopping distance of the $e_{g}$ electrons in the
in-plane or out-of-plane direction, $S$ is the effective spin of a
Mn ion, the Fermi energy $E_{F}$ is measured from the band center,
and $M$ and $m$ are the effective masses of a hole and a spin
wave, respectively. In terms of the hole concentration per unit cell
$n$, the effective hopping integral $t^{\star}$, and the
average spin stiffness $D^{\star}$, the coefficient $B$ can be
rewritten as \cite{zhao2}
\begin{equation}
B=\frac{R \hbar}{48^{2}\pi ^{7} e^{2} S^{2}}\frac{(6\pi ^{2}
n)^{5/3}}{(0.5^{2/3}-n^{2/3})^{9/2}}
(\frac{R^{2}
k_{B}}{D^{\star}})^{9/2}(2.52+0.0017\frac{D^{\star}}{R^{2}
t^{\star}}).
\end{equation}
Here the following simple relationships had been used: $R
k_{F}=(6\pi ^{2} n)^{1/3}$, $M/m=D^{\star}/(R ^{2} t^{\star})$, and
$E_{F}=t^{\star}(6\pi ^{2})^{2/3}(0.5^{2/3}-n^{2/3})$.

\begin{figure}[t]
\begin{center}
\includegraphics[width=\columnwidth]{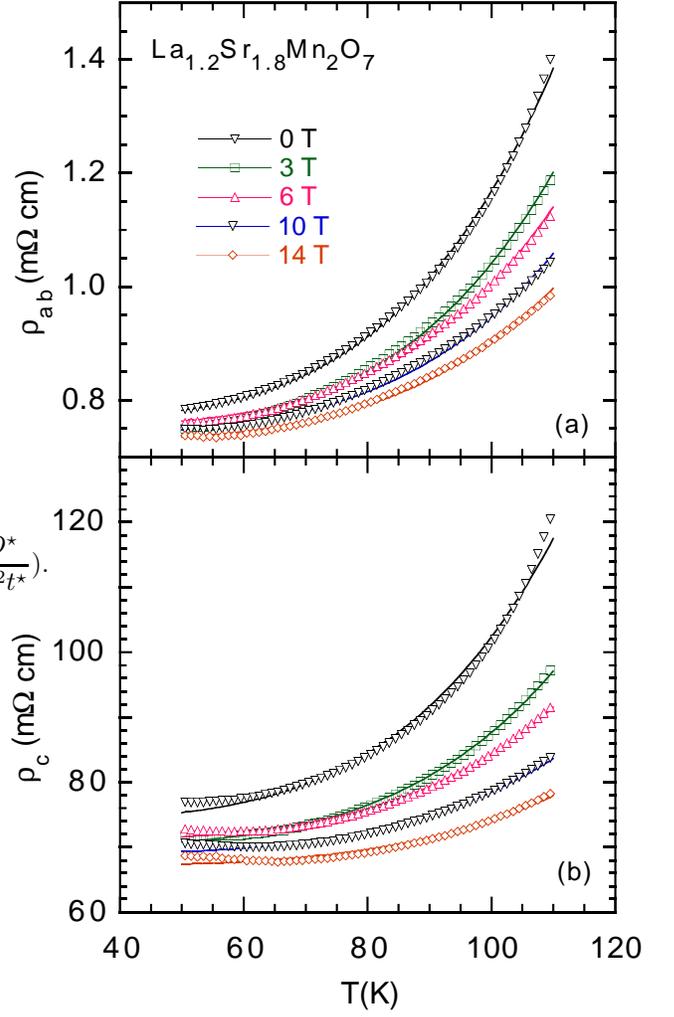}
\end{center}
\caption{ (a) In-plane resistivity $\rho_{ab}$ and (b) out-of-plane
resistivity $\rho_{c}$ vs temperature $T$ for
La$_{1.2}$Sr$_{1.8}$Mn$_{2}$O$_{7}$ measured under various applied magnetic
fields $H$.  The solid lines are fits of the data with a $T^{9/2}$ dependence.}
\end{figure}

To reveal the two-magnon scattering nature of the resistivity in
La$_{1.2}$Sr$_{1.8}$Mn$_{2}$O$_{7}$, we plot in Figs. 3(a) and 3(b) 
the measured
$\rho_{ab}(T)$ and $\rho_{c}(T)$, respectively, along with the fits 
of these data
with $\rho (T)=\rho_0 +BT^{9/2}$. Both $\rho_{ab}(T)$
and $\rho_{c}(T)$ follow remarkably well a $T^{9/2}$ dependence
(solid lines in the figures) in the $T$ range of 50 to 110 K for
different applied magnetic fields. The fitting parameters for
$\rho_{ab}(T)$ and $\rho_{c}(T)$ measured in zero $H$ are
$B_{ab}^{fit}=4.04 \times 10^{-13}$ $\Omega$ cm/K$^{9/2}$ and
$B_{c}^{fit}=2.83\times  10^{-11}$ $\Omega$ cm/K$^{9/2}$, respectively.

\begin{figure}[t]
\begin{center}
\includegraphics[width=\columnwidth]{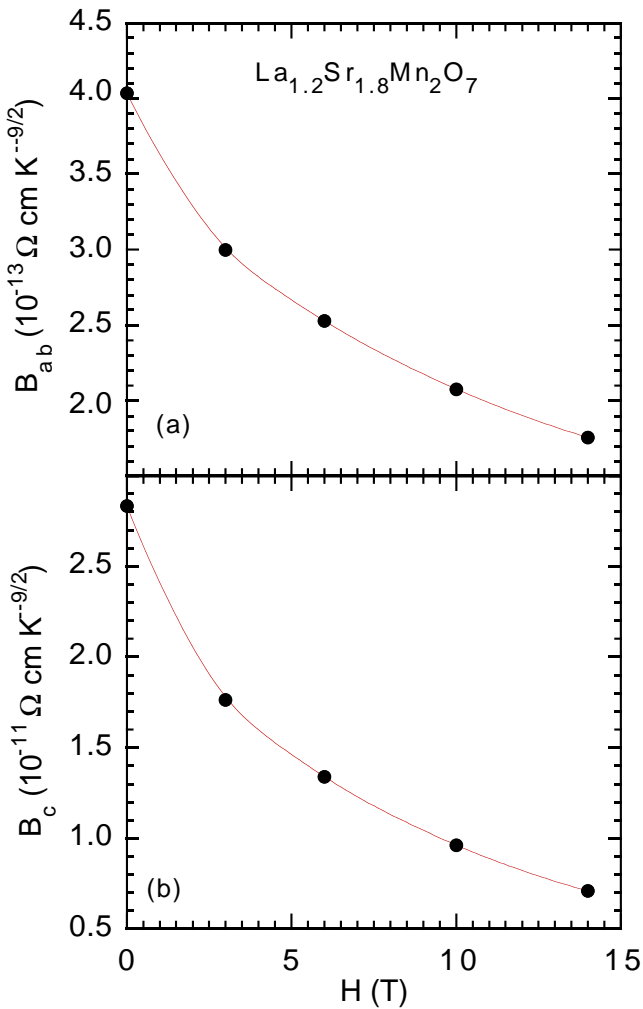}
\end{center}
\caption{Magnetic field $H$ dependence of the coefficients (a) $B_{ab}$ and
(b) $B_{c}$ of the $T^{9/2}$ functional dependence
of the resistivities in the temperature
range 50 to 110 K for La$_{1.2}$Sr$_{1.8}$Mn$_{2}$O$_{7}$. The lines are guides
to the eyes.}
\end{figure}

We determine next the zero-field value of $B_{ab}$ from Eq. (2) and
compare it with the value of the corresponding fitting parameter. In
La$_{1.2}$Sr$_{1.8}$Mn$_{2}$O$_{7}$, the hole concentration $n\equiv
x=$ 0.40, the in-plane hopping distance is the Mn-Mn distance
$R_{ab}= 3.87 \AA$,\cite{mitc} the magnitude of the effective spin is
$S=1.8$,\cite{chat} the effective stiffness constant
$D_{ab}^{\star}=$ 151 meV $\AA ^{2}$ based on recent neutron
scattering measurements,\cite{hf} and $t^{\star}=$ 40 meV as estimated
from the measured effective plasma frequency.\cite{zhao2} With these
values for different physical quantities, Eq. (2) gives
$B_{ab}=1.01\times 10^{-13}$ $\Omega$ cm/K$^{9/2}$. This value of
$B_{ab}$ has the same order of magnitude as the fitting parameter
$B_{ab}^{fit}$, showing that the two-magnon scattering can account
for the $T$ dependence of resistivity of
La$_{1.2}$Sr$_{1.8}$Mn$_{2}$O$_{7}$ in the metallic range of temperatures.

In a conventional Heisenberg system, the spin-wave stiffness $D$
scales with the strength of the magnetic exchange coupling $J$ and
can be expressed as $D=JSR^{2}$.\cite{khal} Since $0.0017
D^{\star}/R^{2} t^{\star} \ll 2.52$, one can ignore this term from
Eq. (2). Hence, one obtains the following expression for the ratio of
the in-plane $J_{ab}$ and interlayer $J_{c}$ exchange
interactions:
\begin{equation}
\frac{J_{ab}}{J_{c}}=(\frac{B_{c}}{B_{ab}}\frac{R_{ab}}{R_{c}})^{2/9}.
\end{equation}
With the values of $B_{ab}$
and $B_{c}$ obtained from the fitting, the hopping distance $R_{ab}$ given
above, and the out-of-plane hopping distance as the Mn-Mn distance between the
MnO$_{2}$ layers $R_{c}=3.88$ $\AA$,\cite{mitc} the above equation gives
$J_{ab}/J_{c}=$ 2.6. This value is in excellent agreement with the value of 2.8
determined  from inelastic neutron scattering measurements on
La$_{1.2}$Sr$_{1.8}$Mn$_{2}$O$_{7}$.\cite{chat2} This result further  indicates
that the two-magnon scattering plays a dominant role in both $\rho_{ab}(T)$ and
$\rho_{c}(T)$ for 50 $\leq T \leq$ 110 K.

The field dependences of $B_{ab}^{fit}$ and $B_{c}^{fit}$ of
La$_{1.2}$Sr$_{1.8}$Mn$_{2}$O$_{7}$ for 50 $\leq T \leq$ 110
K are shown in Figs. 4(a) and 4(b), respectively. Both $B_{ab}^{fit}$
and $B_{c}^{fit}$ are $H$ dependent, implying a strong
sensitivity of the two-magnon scattering to external fields, and
saturate at high fields. The decrease of $B_{ab}^{fit}$ and
$B_{c}^{fit}$ with increasing $H$ is the source of the small negative
magnetoresistance for 50 $\leq T \leq$ 110 K. Their $H$
dependence is also consistent with the spin wave scattering
mechanism. Indeed, the spin-wave scattering itself should decrease
with
increasing $H$. The effect of an applied field is to open an energy
gap $\Delta=g\mu (H+4\pi M_{s})$ in the magnon spectrum. This
can be argued simply on the basis of a reduction in the spin-wave
density by the applied field due to an increase in the energy gap
appearing in the dispersion relation for the magnon energy,
$\varepsilon_{p}=Dq^{2}+\Delta$, where $q$ is the magnon-wave vector.

\begin{figure}[t]
\begin{center}
\includegraphics[width=\columnwidth]{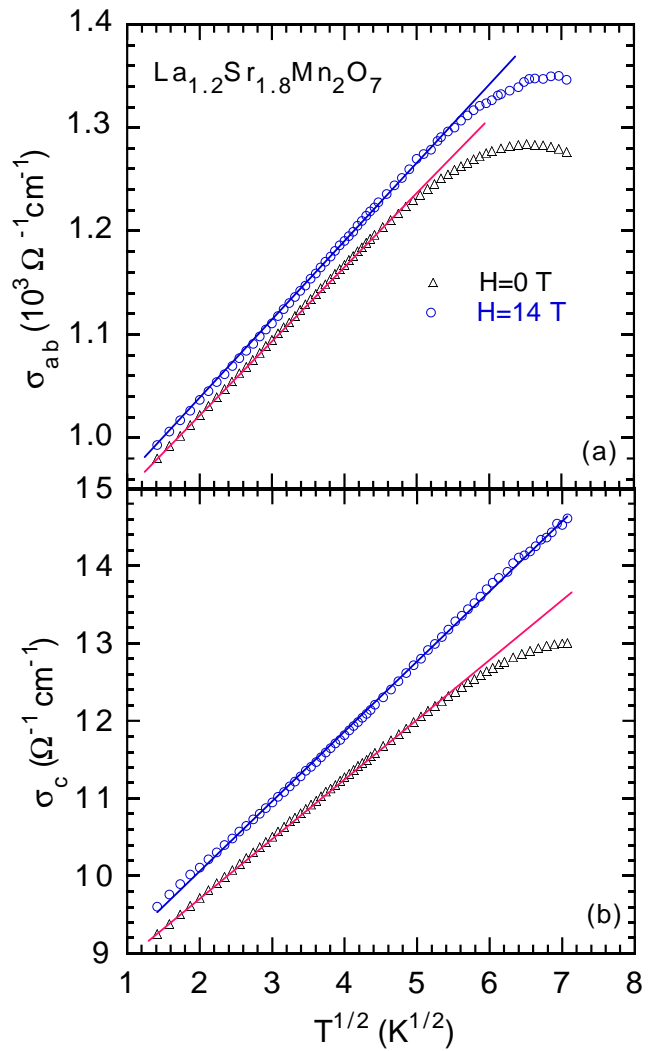}
\end{center}
\caption{(a) In-plane conductivity $\sigma_{ab}$ and (b) out-of-plane
conductivity $\sigma_{c}$ vs $T^{1/2}$ for
La$_{1.2}$Sr$_{1.8}$Mn$_{2}$O$_{7}$ measured down to 1.9 K and in applied
magnetic fields $H=$ 0 and 14 T.}
\end{figure}

Figures 5(a) and 5(b) are plots of $\sigma_{ab}(T)$ and
$\sigma_{c}(T)$, respectively, of La$_{1.2}$Sr$_{1.8}$Mn$_{2}$O$_{7}$
in
the low-$T$ range, down to 1.9 K, measured in $H=$ 0 and 14 T. Both
$\sigma_{ab}$ and $\sigma_{c}$ follow a $T^{1/2}$ dependence below a certain
$T$, which increases with increasing $H$.  This is the first report of a
$T^{1/2}$ dependence of
$\sigma_{c}$ of a bilayer manganite. The $T^{1/2}$ dependence of
$\sigma_{ab}$ is in agreement with previous reports both
in La$_{1.2}$Sr$_{1.8}$Mn$_{2}$O$_{7}$ \cite{chun} and in
La$_{1.3}$Sr$_{1.7}$Mn$_{2}$O$_{7}$ for 30 mK $\leq T \leq $ 2
K.\cite{okuda} It has been suggested \cite{okuda} that the observed
$T^{1/2}$ dependence of $\sigma_{ab}$ is consistent with weak
localization effects in ordered 3D metals,\cite{lee,alts} where the
density of states at the Fermi level has a $T^{1/2}$
singularity due to the influence of interference of the inelastic
electron-electron interaction and the elastic impurity scattering
of the electrons.\cite{alts} In the present case, the electrons are
diffusive instead of freely propagating, leading to a profound
modification of the traditional view based on the Fermi-liquid theory
of metals. Considering the strong anisotropy of the crystal
structure and of the conductivity, the bilayer manganites appear to
be more 2D like. However, $\sigma_{c}(T)$ clearly follows the
same $T^{1/2}$ dependence at low temperatures, consistent with 3D
weak localization effects in disordered metals. This $T^{1/2}$
behavior in {\it both} $\sigma_{ab}(T)$ and $\sigma_{c}(T)$ confirms the 3D
nature of the metallic state and is in good agreement with the
theory of quantum interference in highly anisotropic layered metals
developed by Abrikosov.\cite{abri}

\section{Conclusion}
We performed simultaneous in-plane and out-of-plane resistivity
measurements on bilayer manganite
La$_{1.2}$Sr$_{1.8}$Mn$_{2}$O$_{7}$ single crystal in magnetic fields
applied parallel to the $ab$-plane. Both $\rho_{ab}(T)$ and
$\rho_{c}(T)$ display a $T^{9/2}$ dependence for 50 $\leq T \leq$ 110
K. This $T$ dependence and the magnitude of the fitting
parameter are consistent with two-magnon scattering mechanism. The 
excellent agreement
between the ratio of the exchange interactions $J_{ab}/J_c$ obtained by fitting
the data and
that deduced from neutron scattering provides further support for the 
validity of this
scattering mechanism. Below  50 K, both $\sigma_{ab}(T)$ and 
$\sigma_{c}(T)$ follow a
$T^{1/2}$ dependence, which is consistent with the theory of quantum
interference or weak-localization effects in 3D disordered
metals. The same temperature dependence for both conductivities
strongly indicates that the bilayer
La$_{1.2}$Sr$_{1.8}$Mn$_{2}$O$_{7}$ has a 3D metallic nature at $T<T_c$.

\begin{acknowledgments}
This research was supported  at KSU by the National Science
Foundation under Grant No. DMR-0102415. Work at LANL was performed 
under the auspices of the U. S.
Department of Energy.
\end{acknowledgments}

\end{document}